# Hybrid Job scheduling Algorithm for Cloud computing Environment

Saeed Javanmardi[1], Mohammad Shojafar[2], Danilo Amendola[2], Nicola Cordeschi[2], Hongbo Liu[3], and Ajith Abraham[4,5]

[1]Department of Computer Engineering, Dezful branch, Islamic Azad University, Dezful, Iran
`saeedjavanmardi@gmail.com`
[2]Department of Information Engineering Electronics and Telecommunications (DIET),
University Sapienza of Rome, Rome, Italy {`Shojafar, Cordeschi, Amendola`}`@diet.uniroma1.it`
[3]School of Information, Dalian Maritime University, 116026 Dalian, China `hb@dlut.edu.cn`
[4]IT4Innovations, VSB-Technical University of Ostrava, Czech Republic
[5]Machine Intelligence Research Labs (MIR Labs), Scientific Network for Innovation and Research Excellence, Auburn, USA
`ajith.abraham@ieee.org`

**Abstract.** In this paper with the aid of genetic algorithm and fuzzy theory, we present a hybrid job scheduling approach, which considers the load balancing of the system and reduces total execution time and execution cost. We try to modify the standard Genetic algorithm and to reduce the iteration of creating population with the aid of fuzzy theory. The main goal of this research is to assign the jobs to the resources with considering the VM MIPS and length of jobs. The new algorithm assigns the jobs to the resources with considering the job length and resources capacities. We evaluate the performance of our approach with some famous cloud scheduling models. The results of the experiments show the efficiency of the proposed approach in term of execution time, execution cost and average Degree of Imbalance (DI).

## 1 Introduction

Cloud computing is composed of distributed computing, grid computing, utility computing, and autonomic computing [1]. In cloud computing, users do not know where the services located in which part of the infrastructure. The users only use the services through the cloud infrastructure paradigm and pay for the requested services [2]. Cloud infrastructure provides on demand access to some shared resources and services. These services are provided as a service over a network, and can be accessed over the internet [3].

Scheduling algorithms [4] are used mainly to minimize the execution time and execution cost. Scheduling, handles the problem of which resources needed to be assigned for the received job. A good scheduling algorithm should consider the load balancing of the system and total execution time of the available resources. In one hand, it should reduce the execution time and from the other hand, it should reduce the execution time. For achieving both of them it is better not to waste the resources with high capacities to the jobs with low length. The scheduler should assign the jobs to the resources according to the job length and resources capacities [5, 6].

Recently, a lot of attention has been paid to the usage of intelligent approaches in cloud job scheduling [7, 8]. Genetic algorithm [9, 10] and fuzzy theory [11] are two famous artificial intelligence approaches which are used in this paper. Genetic algorithm starts with a set of chromosomes called population. Then with the usage of a fitness function the fitness value of the chromosomes are calculated. After that the best two chromosomes are selected and then the crossover operation is done. After that, standard genetic algorithm performs the mutation operation. It mutates the new child at some positions. Finally the algorithm adds the new chromosome to the population. It will be continued until the termination condition happens. Fuzzy theory is a logic which is less severe than the computation computers usually perform. In recent years fuzzy logic has been used in distributed systems like grid and cloud not only for scheduling, but also for trust management and resource discovery [12]. Fuzzy Logic tenders various singular characterizes that make it a particularly good optional for many control problems such as [13]. Fuzzy Logic manages the examination of knowledge by utilizing fuzzy sets, each of which can show a linguistic phrase such as "Bad", "Medium", etc. [14].

In the proposed approach with the aid of fuzzy theory, we try to modify the genetic algorithm; we use fuzzy system in fitness step and cross over step. The goal of using fuzzy theory in genetic algorithm is to reduce the iteration of producing the population and assigning the suitable resources to the jobs based in the node capacities and length of the jobs. The new algorithm obtains the best chromosomes in a few iterations.

The rest of this paper is as follows: in the next section we provide related works; in this Section, we take a brief look at some works, which are about cloud job scheduling. In Section 3 we describe our approach. The performance evaluation and experimental results are presented in Section 4. Finally, in Section 5, we make a conclusion.

## 2 Related Works

Load balancing is a main concept in large scale computing such as cloud whose aim is to guarantee that every computing resource is distributed efficiently [15]. The main goal of cloud computing is to assign the jobs across a large number of distributed resources. In recent years, a lot of attention has been paid to the artificial intelligence methods such as genetic algorithms and fuzzy theory by researchers because of its intelligence and inferred parallelism [16]. Genetic algorithm has been extremely usage to solve the problem of cloud resources scheduling and has obtain perfect effects [17]. Specifically, authors in [17] with the use of Genetic algorithm proposed a cloud scheduling approach for VM load balancing. As the authors mentioned system reliability and availability are some features in cloud that should be considered; the authors claim their approach effectively improve overall system reliability and availability.

Authors in [18], describe several job scheduling algorithm and compare between these algorithms. As it is mentioned in this paper, a good cloud job scheduling algorithm should schedule the resources to optimize the usage of the resource. Various scheduling algorithms are presented for resource scheduling but each one has its own restriction. These algorithms represent optimum or non-optimum solution for the problems. At this time, we require more exact algorithm for resource scheduling which is the major research challenge. In case of Ant Colony Optimization (ACO) when more resources are engaged, ACO produce colonies thus an ant follows less seemingly pheromone trail from another colony. By chance that Particle Swarm Optimization the solution space or search space can be very large.

Authors in [19] proposed a job-oriented based model for cloud resource scheduling. This model assigns jobs to the resources according to the rank of the job. This paper also discusses the analysis of resource scheduling algorithm such as time parameters of Round Robin, Pre-emptive Priority and Shortest Remaining Time First.

In [20], authors proposed a model to deal with the job scheduling problems for a group of cloud user requests. Each datacenter has different services with various resources. This plan assumes resource provisioning as an important issue for job scheduling. The main goal of this model is reducing the average tardiness of connection requests. This paper present four merged scheduling algorithms and used to schedule virtual machine on data centers. Of the four methods, the method merging Resource Based Distribution technique and Duration Priority technique have represent the best performance becoming the minimum tardiness while consent to the problem constraints. The mentioned model reduces the average tardiness of connection requests and the connection blocking percentage.

In [21], proposed a Genetic algorithm based job scheduling in which there is a fitness function which divided into three sub-fitness function and then linear combination of these sub-fitness value is carry out for obtaining the fitness value. This paper uses a pre-migration strategy which is based on three load dimension: CPU utilization, network throughput, disk I/O rate. TO achieve a nearly optimum solution this plan applies the hybrid genetic algorithm merge with knapsack problem with multiple fitness. The author claims that the algorithm can obtain the goal of raising resources utilization efficiency and lower energy consumption. The algorithm reduces energy consumption and also increases the utilization of the resources.

Reference [22] proposes a genetic based job scheduling approach to load balance the virtual machines in a large scale cloud infrastructure. The author claims that his plan solves the problem of load imbalance and high migration costs. This approach modifies the standard genetic algorithm to obtain the mentioned results. This approach has six steps and stopping condition for the algorithm is if there exists a tree which meets the heat limit requirement.

## 3  Proposed Approach

In the proposed approach, a job is represented as a gene, which should be assigned to the computational resources; and a set of genes create a chromosome.

In fact, gene is defined as a job, which should be assigned to computational resources. We create two types of chromosomes based on different criteria. The first type is created based on the job length, CPU speed of the resources and ram value of the resources. The second type is created based on the job length and the bandwidth of the resources. These criteria are the input parameters of the fuzzy system. In each type of chromosomes, some random populations of chromosomes (a set of jobs) are created and represented as n chromosomes; then the computational resources are assigned to the chromosomes randomly and then the algorithm calculates the fitness value of every chromosome of each type of chromosomes. The fitness value is achieved by fuzzy system. Then, the algorithm selects two chromosomes individuals from the mentioned two types of chromosomes according to the fitness value.

Then the algorithm performs the crossover operation with the aid of fuzzy theory for these two chromosomes. At the end of this step, a new chromosome will be created which is the best chromosome of the first generation. Just like to the other genetic based cloud scheduling approaches, we add the obtained chromosome to the previous population and we use the new generation as the current generation. The algorithm repeats till the selected two chromosomes for cross over step are homolog. The detail of the proposed approach is as follow.

The main purpose of our algorithm is assigning the most suitable resources to the jobs based on the bandwidth and computational capacities of resources and the job length. The algorithm tries to assign the jobs with high length to the resources with high bandwidth and high computational resources. We use fuzzy theory in two steps of our algorithm. First, it is used for calculating the fitness value F(x) of every chromosome in the mentioned two types of chromosomes. Then it is used in crossover step of our algorithm. We do not use the typical crossover approaches such as single-point crossover or two-point crossover. We use fuzzy theory to cross over the two chromosomes genes; the job length of the chromosomes genes, VM bandwidth, VM mips and the amount of ram of the assigned resources to the chromosomes genes are the input parameters of the fuzzy system in this step. With the aid of fuzzy theory the selection of chromosomes and assigning the resources to the jobs will be targeted. There are two common types of fuzzy inference systems; Mamdani and Sugeno [14]. Mamdani [14] inference system is used in our approach due to its simplicity and entitlement of defining rules based on the last experiences. With the aid of fuzzy rules we define the priority of the input parameters of the algorithm. For example, consider this rule: If job length is high and bandwidth is low and the amount of Ram is medium and CPU speed is high then the result is adequate. In this rule we define the priority of the input parameters. In this rule for a job with high length, the priority of CPU speed is high and the priority of Ram is medium and the priority of bandwidth is low.

Formally speaking, at first, the algorithm produces random population of two types of chromosomes. The first type is created based on the job length, CPU speed of the resources and Ram value of the resources. The second type is created based on the job length and the bandwidth of the resources. Then the algorithm calculates the fitness value F(x) of genes of the chromosomes in the mentioned two types of population. Calculation of the fitness value is with the aid of fuzzy theory. The mentioned parameters are the inputs of fuzzy inference system which is used for fuzzy reasoning. The output value of fuzzy inference system is a non-fuzzy number which determines the fitness value of each chromosome. Fig. 1 represents the used fuzzy inference system.

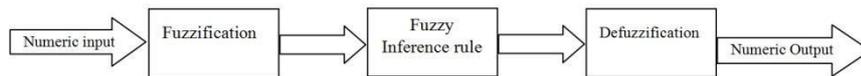

**Fig. 1.** The structure of the fuzzy inference engine

For calculating the fitness value first, fuzzy inference system receives the input parameters and determines the degree to which they belong to each of the suitable fuzzy sets through membership functions. To perform this, three overlapping fuzzy sets are created. It is better to determine the intervals in a way that the endpoint of the first fuzzy set be the starting point of the third fuzzy set. A or *membership function* is a curve that defines how each point in the input space is mapped to a membership degree between 0 and 1 [23]. $\mu$ shows the membership degree which is a number between 0 and 1. Generally, we have the following equation (1).

$$\mu_A(x) = Degree(x) \text{ in } A \qquad (1)$$
$$\forall x \in X : \mu_A(x) : X \to [0,1]$$

Figs. 2, 3 show the fuzzy sets for length of job, VM mips (or CPU power) parameters which are created by using MATLAB fuzzy logic toolbox [24]. These fuzzy sets are used for the first and second experiments of performance evaluation step.

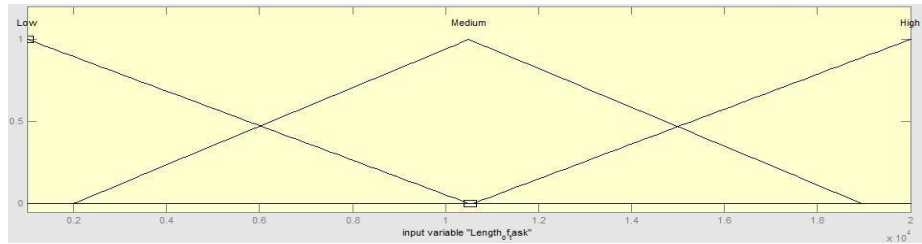

**Fig. 2.** Fuzzy sets for length of jobs parameter

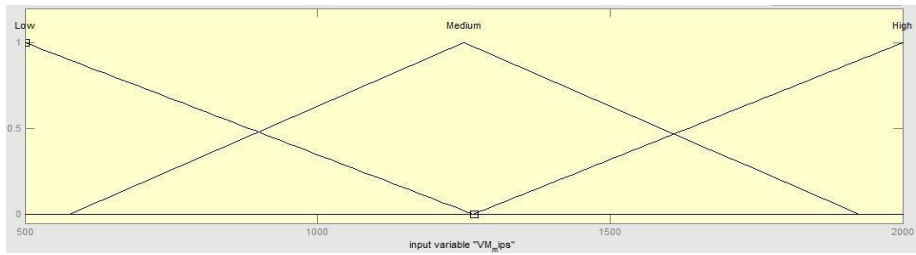

**Fig. 3.** Fuzzy sets for VM mips

For example in Fig. 3, with the VM mips 1000, the membership degree for low interval is 0.3, for medium interval is 0.7 and for high interval is 0. These values are used for fuzzy rules in fuzzy reasoning step. The fuzzy rules of Mamdani [14] inference system in our approach are defined based on the cloud environment and it's administering policy, the general concept of the decision concept explained in TOGA [25].

After selecting the best two chromosomes of each type (the parents), the algorithm performs the crossover operation. In this step, the job length of the genes of the first chromosome is the first parameter of fuzzy system. The VM mips is the second parameter of fuzzy system. The assigned resources to the genes of the chromosomes (the first gen of the first chromosome with the first gen of the second chromosome, the second gen of the first chromosome with the second gen of the second chromosome and so on) are exchanged. The algorithm repeats till stopping condition is satisfied. Stopping condition is being the selected two chromosomes for cross over homolog. The output chromosome of crossover operation has the genes, which include the jobs assigned to the most suitable resources. Due to using fuzzy theory in fitness step and cross over step the algorithm achieves the best chromosome in a few iterations. In fact, the genes of the mentioned chromosome assign the most suitable resources to the jobs in order to reduce makespan and cost.

adfa, p. 6, 2014.
© Springer-Verlag Berlin Heidelberg 2014

## 4 Performance evaluation

The performance evaluations of the proposed approach and the comparison with other algorithms have been implemented on the CloudSim. For the first and second experiments, we compare our approach with ACO and MACO algorithms [26] in terms of Makespan and the degree of imbalance such as in [27, 28]. These two experiments are carried out with 10 Data centers, 50 VMs and (100-1000) cloudlets (jobs) under the simulation platform. The job length is from 1000 MI (Million Instructions) to 20000 MI. Table 1 represents parameter settings of the first and second experiments.

Table 1. Parameter settings for the first and second experiments

| Parameters | Values |
| --- | --- |
| Length of job | 1000-20000 |
| Total number of jobs | 100-1000 |
| Total number of VMs | 50 |
| VM mips | 500-2000 |
| VM memory (RAM) | 256-2048 |
| VM Bandwidth | 500-1000 |
| Number of PEs requirements | 1-4 |
| Number of datacenters | 10 |
| Number of Hosts | 2-6 |

In the first experiment, the performance evaluation is compared in term of the average makespan with different number of jobs. The average makespan of the proposed approach, MACO and ACO algorithms are presented in Fig. 4. In Fig. 4, while the numbers of jobs are increased, the makespan increased, but the increment ratio in our approach is much lower than ACO and MACO, because, our system tries to find the optimum scheduler based on joint of local and global searchers.

For the second experiment, we compare the average Degree of Imbalance (DI) which represents the imbalance among VMs. DI is calculated by the following equation (2).

$$DI = \frac{T_{max} + T_{min}}{T_{avg}}, \tag{2}$$

where $T_{max}$, $T_{min}$ and $T_{avg}$ are the maximum, minimum and average total execution time (in seconds) of all VMs respectively, so DI is dimensionless. As it is illustrate in the following figure, new approach has better DI than ACO and MACO.

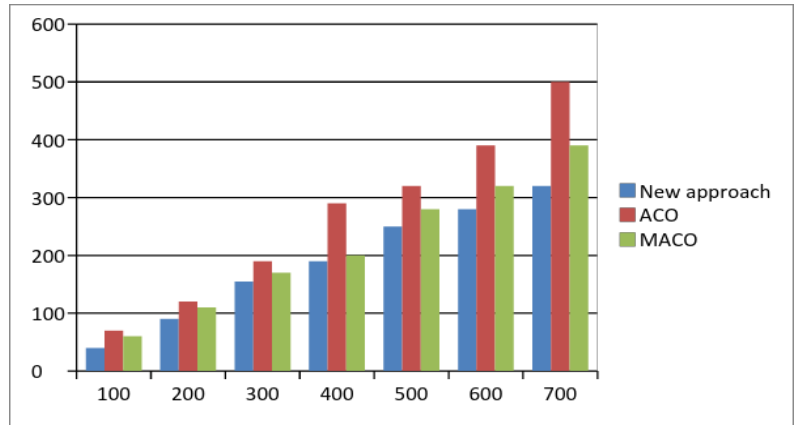

**Fig. 4.** Average makespan

It is because of assigning the resources to the jobs according to the jobs length. The new algorithm assigns jobs with the higher size to the powerful resources. In another word, the new algorithm assigns jobs to the resources with considering job length and resource ability. So, total execution time of each VM will be decreased. That is exactly why new algorithm has better DI. Fig. 5 represents average Degree of Imbalance (DI).

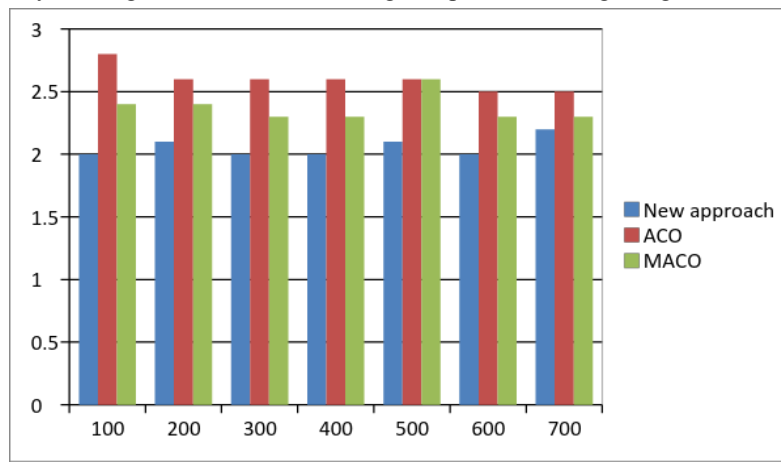

**Fig. 5.** Average Degree of Imbalance (DI)

## 5 Conclusion

In this paper we used genetic algorithm as the basis of our approach and we modify it with the aid of fuzzy theory to reduce the iteration of producing the population. We define two types of chromosomes with different QOS parameters; then with the aid of fuzzy theory we obtain the fitness value of all chromosomes for the mentioned two types. The new approach with the use of fuzzy theory modifies the standard genetic algorithm and improves system performance in terms of execution cost about 45% and total execution time about 50% which are the main goal of this research.

## Acknowledgements

This work was supported in the framework of the IT4 Innovations Centre of Excellence project, reg. no. CZ.1.05/1.1.00/02.0070 by operational programme 'Research and Development for Innovations' funded by the Structural Funds of the European Union and state budget of the Czech Republic, EU.